\documentclass{article}
\usepackage[final]{graphics}
\usepackage{graphicx}
\usepackage{makeidx}
\usepackage{bm}
\usepackage{authblk} 
\usepackage{epstopdf}
\usepackage{float}
\usepackage{subfig}
\usepackage{indentfirst} 
\usepackage{amsmath}
\usepackage{wrapfig}
\usepackage{color}
\usepackage{natbib}
\usepackage{booktabs}
\usepackage{caption}
\usepackage[colorlinks=true,citecolor=blue,linkcolor=blue]{hyperref}
\usepackage[noabbrev,capitalize,nameinlink]{cleveref}

\setcitestyle{authoryear,open={},close={}}
\usepackage[applemac]{inputenc}
\usepackage[portrait]{geometry}

\begin{document}

\title{Fully Resolved Numerical Simulations of Fused Deposition Modeling. Part II---Solidification, Residual Stresses, and Modeling of the Nozzle \\
{\small In preparation for: Rapid Prototyping Journal}}

\author{Huanxiong Xia, Jiacai Lu, and Gretar Tryggvason}
\affil{Department of Mechanical Engineering \\ Johns Hopkins University, MD,  USA}
\date{}

\maketitle
%\end{document}

\begin{abstract}

\noindent{\bf Purpose --- }This paper continues the development of a comprehensive methodology for fully resolved numerical simulations of fusion deposition modeling.

\noindent{\bf Design/methodology/approach --- }A front-tracking/finite volume method introduced in Part I to simulate the heat transfer and fluid dynamics of the deposition of a polymer filament on a fixed bed is extended by adding an improved model for the injection nozzle,  including the shrinkage of the polymer as it cools down, and accounting for stresses in the solid.

\noindent{\bf Findings --- }The accuracy and convergence properties of the new method are tested by grid refinement and the method is shown to produce convergent solutions for the shape of the filament, the temperature distribution, the shrinkage and the solid stresses.

\noindent{\bf Research limitations/implications --- }The method presented in the paper focuses on modeling the fluid flow, the cooling and solidification, as well as volume changes and residual stresses, using a relatively simple viscoelastic constitutive model. More complex material models, depending, for example, on the evolution of the configuration tensor, are not included.

\noindent{\bf Practical implications --- }The ability to carry out fully resolved numerical simulations of the fusion deposition process is expected to be critical for the validation of mathematical models for the material behavior, to help explore new deposition strategies, and to provide the ``ground truth'' for the development of reduced order models.

\noindent{\bf Originality/value --- }The paper completes the development of the first numerical method for fully resolved simulation of fusion filament modeling.
\end{abstract}

\section{Introduction}

As the role of additive manufacturing grows and an increasing range of products are fabricated using the various processes available, demands for reliability are likely to increase. Modeling, in particular, is likely to play a significant role. One of the most common rapid prototyping processes is Fused Deposition Modeling (FDM), or Fused Filament Fabrication (FFF), where a polymer filament is heated and deposited from a moving nozzle in a controlled way, forming layers of a specific shape. Once a layer has been completed, the nozzle is raised, or the platform is lowered, and another layer is laid down, thus building an artifact layer-by-layer. The underlying physics behind FDM/FFF is relatively straight forward and a large number of low cost 3D printers are now available. Although 3D printers that use polymers and build artifacts one layer at a time are most common, the process has been used for a wide variety of materials, including concrete (\cite{Gosselin2016}) and biomaterials (\cite{ChiaWu2015,Weietal2016}). In addition, although building objects layer-by-layer is most common, a large number of ``3D pens'', where the deposition is not limited to specific layers, are also available. While such free-form fabrication devices seem to have significant potential, most 3D pens currently available appear to be intended for hobbyists and artists, rather than for engineering prototyping.

A detailed mathematical model of additive manufacturing processes offers several opportunities, as in most other engineering situations.
First of all, fully resolved simulations are important for modeling and the assessment of the appropriateness and fidelity of the various material models, since it is necessary to solve the governing equations before the results can be compared with experiments. A fully tested mathematical model can be used to explore how sensitive the final outcome is to the inclusion of various physical phenomena and how they are represented. Secondly, the effect of the various operating parameters can be explored and the result of changing them predicted. In addition, a comprehensive mathematical model, even if it is computationally intensive and its use is impractical as a routine design tool, can produce the ``ground truth'' for reduced order models intended for routine design work.

In \cite{Xiaetal:RPJ:2017} a numerical method for fully resolved simulations of the deposition of a filament of hot polymer onto a substrate and its cooling down was introduced. Viscosity was taken to be a strong function of temperature so the object was essentially rigid, once it had cooled down. The method is based on the finite volume/front tracking method originally introduced by \cite{UnverdiTryggvason:JCP:92}, where the conservation equations for mass, momentum and energy are solved on a fixed structured and staggered grid discretizing a domain containing two or more fluids, and the interface between the different fluids is tracked using connected marker particles. The method was extended to FDM/FFF by making it implicit to handle realistic material parameters, the viscosity of the melt was taken to be a function of temperature and shear rate, and a volume source was incorporated to model the nozzle. The method was implemented for a rectangular domain, initially containing air, and the behavior of both the injected polymer and the air is simulated. The performance and accuracy of the method was tested in several ways using simple injection of two short filaments, one on top of the other, and it was found that for governing parameters similar to those encountered for realistic conditions, a converged solution could be obtained using about thirty control volumes across the filament diameter. A more complex geometry, consisting of a two layer infilled rectangular object was also simulated to show the capability of the method. The computational setup was, however, simplified in several ways. The most significant one was that  the injection nozzle was modeled using a simple volume source, and that except for the change in viscosity with temperature no model was included for the material behavior of the polymer as it solidified. 

The only other effort to produce fully resolved simulations of FDM/FFF, known to the authors, was described in \cite{Bellini2002}, where the fluid flow and heat transfer during the extraction of ceramic filaments was simulated for a two-dimensional geometry. Considerable efforts have, however, been devoted to modeling various aspect of FDM/FFF, usually in isolation from other processes. Such studies include \cite{JiZhou2010} and \cite{ZhangChou2008}, where finite element simulations are used to examine the cooling down after the filament has been deposited, and the distortion of FDM parts, respectively. Modeling of the mechanical properties of the final part can be found in 
\cite{BelliniGuceri2003,Ziemianetal2012a, Singamnenietal2012, Domingo-Espinetal2015, Shahrainetal2016, Karamoozetal2014, Naghiehetal2016} and \cite{Lienekeetal2016}, and  \cite{Rezaieetal2013} applied topology optimization to FDM build parts.
The roughness of the final part was examined by \cite{Ahn2009, BoschettoBottini2013}, and the bond between filaments by \cite{Bellehumeuretal2004, Sunetal2008} and \cite{Costaetal2017}.
The flow in the nozzle was studied by \cite{Ramanathetal2008}. \cite{McIlroyOlmsted2017} studied how the 
high strain rate as the polymer flows out of the nozzle can modify the properties of the material and the bonding, and \cite{McIlroyOlmsted2017-2} examined the effect of disentanglement of the polymer in the nozzle on the weld strength between adjacent filaments.
For a review focused on modeling of FDM/FFF see, for example, \cite{Turneretal2014} and \cite{Turneretal2015}. 

%Construction of scaffolds for tissue engineering  are reviewed by \cite{Abdelaal2011}
%\cite{Espalinetal2010} patient specific; \cite{Choietal2011} mobile system

While fully resolved numerical simulation of the type presented here are still very rare for FDM/FFF, efforts to simulate other additive manufacturing processes are starting. For laser powder bed fusion processes, where a laser is used to selectively fuse or sinter metal particles, researchers at the Lawrence Livermore National Laboratory have developed comprehensive computational models  that have been used to examine the denudation and other aspects of metal powders, see \cite{KhairallahAnderson2014,Kingetal2015-2,Kingetal2015,Markletal2015, Matthewsetal2016} and \cite{Khairallahetal2016}. Several years ago there was also considerable interest in examining how objects could be built either by depositing drops one by one (\cite{GaoSonin:94}) or by spraying molten metal on a substrate (\cite{Liuetal:93,ChungRangel:01,Mostaghimietal:02,Cheetal:JAMM:2004}). For recent reviews of additive manufacturing in general, including FDM/FFF, see \cite{Bourell2016, Bikasetal2016} and \cite{Slotwinskietal2016}, for example, as well as  \cite{NAP23646} for a report on a recent workshop on the modeling of additive manufacturing. 

The rest of the paper is laid out as follows. In the next section the extensions of the numerical model introduced in Part I are described. The governing equations are first summarized, the modifications needed to include volume change due to temperature are explained, the computations of solid deformation and stresses are presented and a new model to better describe the injection are introduced. A short description of minor modifications of the flow solver concludes the method section. In the result section the new nozzle model is first tested by grid refinement, followed by similar tests for the solidification, volume change and solid stress computations, using two short filaments deposited on top of each other. The result section concludes with simulations of a more complex shape. Possible further extensions of the methodology are discussed in the conclusion section.

\begin{figure}
\centerline{\scalebox{0.5}{\includegraphics{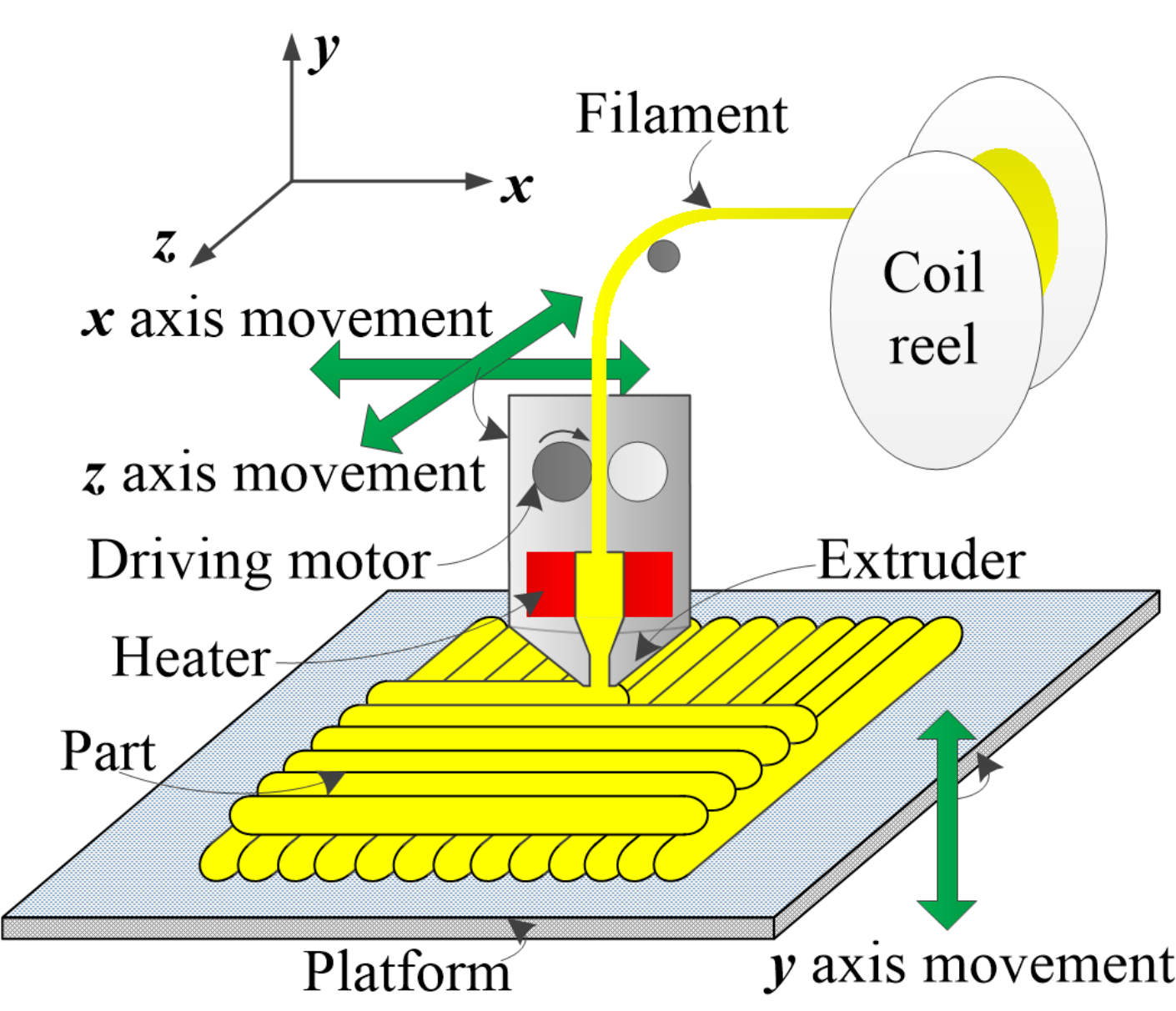}  }}  
\caption{A schematic.}
\label{figure1}
\end{figure}

\section{Formulation and Numerical Method}

For a basic description of the algorithm for the fluid flow and the heat transfer, see Part I (\cite{Xiaetal:RPJ:2017}). Here the focus is on modeling of the properties of the polymer as it solidifies and the nozzle injecting the hot polymer. Figure \ref{figure1} shows the process schematically. The polymer is shown in yellow and the motions of the platform and the nozzle are shown by the green arrows.

\subsection{Governing Equation}

The behavior of the molten and solid polymer and the air is governed by the conservation of mass, momentum and energy and one set of equations is used for the whole computational domain. The mass conservation equation is:
\begin{equation}
{\partial \rho \over \partial t}+\nabla \cdot \rho {\bf u}=\rho \dot Q \delta ({\bf x} -{\bf x}_S).
\label{massconservation}
\end{equation}
Momentum conservation is:
\begin{equation}
{\partial \rho  {\bf u} \over \partial t}+ \nabla \cdot (\rho  {\bf u}{\bf u})=\nabla \cdot {\bm \sigma}  + \tilde \rho {\bf g}+ \sigma \int_F \kappa_f {\bf n}_f \delta ({\bf x} -{\bf x}_f ) dA_f.
\label{navierstokes}
\end{equation}
And the energy equation is: 
\begin{equation}
{\partial \rho c_p T\over \partial t}+\nabla \cdot \rho c_p {\bf u}T =\nabla \cdot k \nabla T+ \rho c_p T_{inj} \dot Q \delta ({\bf x} -{\bf x}_S).
\label{energyconservation}
\end{equation}
Here, ${\bf u}$ is the velocity field, $T$ is the temperature, $T_{inj}$ is the injection temperature, ${\bf g}$ is the gravity acceleration, ${\bm \sigma}$ is the stress field, and $\rho$, $c_p$, and $k$ are the density, heat capacity and thermal conductivity, respectively.  In general the material properties are different in the different fluids and phases, and change discontinuously across the interfaces separating those.  $\dot Q$ is the volumetric rate of polymer injection and $\delta$ is a three-dimensional delta function located at ${\bf x}_f$ or ${\bf x}_S$. The source term in the energy equation keeps the injected material at the desired temperature, but other heat sources, including frictional heating, latent heat due to phase change and radiation, are neglected. To identify the polymer, an index function $I$ that is 0 in the region occupied by the polymer and 1 in the air is used.  The surface tension coefficient for the polymer/air boundary is $\sigma$, and $\kappa_f$ and ${\bf n}$ are the interface curvature and normal vector. To simplify the implementation of the algorithm, surface tension for the air/solid interface, as well as for the air/polymer, is computed since it is much smaller than the solid stress, and including it has no effect on the results. Since part of the polymer has solidified, an additional index function is introduced such that $I_s=1$ if $T  \le T_m$ and zero if $T > T_m$, where $T_m$ is the solidus point. The polymer and the air are assumed to be incompressible, since all velocities are very low, but for the air buoyancy is included using the Boussinesq approximation, as in Part I (in equation (\ref{navierstokes}), the density multiplying the gravity acceleration is therefore denoted by $\tilde \rho$). The stresses in the solid and fluid are found using different mathematical expressions, but can be combined by writing: 
\begin{equation}
{\bm \sigma}=-p{\bf I}+(1-I_s) {\bm \sigma}'_f+I_s {\bm \sigma}'_s,
\label{eqSDt}
\end{equation}
where ${\bf I}$ is the unit tensor, and ${\bf \sigma}'$ stands for the deviatoric part of the stress tensor. The subscripts $s$ and $f$ identify the solid and fluid phase, respectively. 
The deviatoric tensor for the fluid stress can be found from the velocity field, ${\bm \sigma}'_f=2\mu_f {\bf D}$,  where ${\bf D}=(1/2)(\nabla {\bf u}+\nabla {\bf u}^T)$ is the rate of deformation tensor. The computation of the solid stress ${\bm \sigma}'_s$ is more complex and is discussed below. 

The pressure, in both the fluid and the solid, is found by solving a Poisson equation, as described in Part I. Volume changes due to temperature changes, but not due to stresses, are accounted for as discussed in the next section. Thermal stress is given by $-3\kappa \alpha \Delta T {\bf I}$ (\cite{teskeredvzic2002}) for an isotropic material, where $\kappa$ is the bulk modulus (Lam\'e parameter) of the material, $\alpha$ is the thermal expansion coefficient and $\Delta T$ is the temperature change. This expression shows that thermal stress, resulting from volume changes, is isotropic and can therefore be absorbed into the pressure.

\subsection{Volume Change due to Temperature Change}

In general, the density is a function of both pressure and temperature, $\rho = \rho (p,T)$, but for a polymer filament cooling down and solidifying in air under atmospheric conditions, it is assumed that the effect of pressure  can be neglected, so that $\rho = \rho (T)$. 
The conservation of mass equation can be written as
\begin{equation}
{\partial \rho \over \partial t}+{\bf u} \cdot \nabla  \rho +\rho \nabla \cdot {\bf u}=\rho \dot Q \delta ({\bf x} -{\bf x}_S),
\end{equation}
or
\begin{equation}
\nabla \cdot {\bf u}=\dot Q \delta ({\bf x} -{\bf x}_S)-{1 \over \rho} \biggl( 
{\partial \rho \over \partial t}+{\bf u} \cdot \nabla  \rho \biggr)=\dot Q \delta ({\bf x} -{\bf x}_S)-{1 \over \rho} {D \rho \over Dt},
\end{equation}
where $D()/Dt$ is the usual convective derivative. Since the density is a function of temperature only, it is possible to write
$D \rho/Dt = (d \rho / dT) (DT /Dt) = -\rho^2 (d v/ d T) (DT /Dt) $, where $v=1/\rho$ is the specific volume and $ d\rho/dT=-\rho^2 d v / dT$. 
Thus:
\begin{equation}
\nabla \cdot {\bf u}=\dot Q \delta ({\bf x} -{\bf x}_S)+{\rho}\Bigl( {d v \over d T} \Bigr) {D T \over Dt}.
\end{equation}
By using the conservation of mass equation and assuming a constant thermal capacity, the energy equation can be rewritten  as
\begin{equation}
{D T \over Dt}= {1 \over \rho c_p} \Bigl( \nabla \cdot k \nabla T - \rho c_p T_{inj}\dot Q \delta ({\bf x} -{\bf x}_S)\Bigr),
\end{equation}
resulting in
\begin{equation}
\nabla \cdot {\bf u}=\dot Q \delta ({\bf x} -{\bf x}_S)+{1 \over c_p }\Bigl( {d v \over d T} \Bigr) \Bigl( \nabla \cdot k \nabla T  -\rho c_p T_{inj} \dot Q \delta ({\bf x} -{\bf x}_S)\Bigr) .
\label{volumesource}
\end{equation}
Equation (\ref{volumesource}) gives the total volume change, where the first term on the right hand side is due to the material injection and the second term is due to temperature change due to conduction.   The modified equation for the divergence of the velocity leads to a slight change in the pressure equation that was given in Part I, to account for the additional volume change.

\subsection{Solid Deformation and Stress}
Unlike the fluid stresses, the solid stresses are functions of the gradients of the deformation, rather than the deformation rates (velocities). Since the solid stresses are found on the fixed grid, just like the fluid stresses, it is necessary to compute the deformation gradients in an Eularian frame of reference.
To do so, a reference configuration ${\bf X}=(X_1,X_2,X_3)$, which is the location of a material point at the time when the polymer solidifies, is introduced. 
Since a material point moves with the velocity, the evolution of  the coordinates of the reference configuration, $X_j$, is given by:
\begin{equation}
{D X_j \over Dt}=0.
\label{eqDXDt}
\end{equation}
To find the stresses, the deformation gradients, $F_{ij} = \partial x_i /\partial X_j$ are needed. An equation for the inverse, $f_{ji} ={\partial X_j / \partial x_i} $, can be found by differentiating equation (\ref{eqDXDt}):
\begin{equation}
{\partial^2 X_j \over \partial x_i \partial t}+{{\partial  \over \partial x_i} {(u_k{\partial X_j \over \partial x_k})}}=0.
\label{eqDD}
\end{equation}
Using the chain rule equation (\ref{eqDD}) can be rewritten as:
\begin{equation}
{\partial f_{ji} \over \partial t}+u_k {\partial  f_{ji} \over \partial x_k}+{\partial u_k \over \partial x_i} f_{jk} =0.
\end{equation}
To find the solid stresses, the inverse, ${\bf F}={\bf f}^{-1}$, is first found and then the left Cauchy-Green tensor, ${\bf B}={\bf FF}^T$, is computed.
Here, a neo-Hookean elastic solid model is used to describe the solidified polymer. However,  the method can easily be modified to use other stress models. The neo-Hooken solid is a nonlinear stress-strain relationship suitable for predicting the behavior of the polymer as the stress-strain relationship changes with the state of the cross-linked polymer chains. Specifically, with the above hypothesis for the volume change, the solid stress for the solidified material is given by:
\begin{equation}
{{\bm \sigma}_{s,e}}=GJ^{-{5\over3}}{\bf B}'-p{\bf I}={{\bm \sigma}'_{s,e}}-p{\bf I},
\label{eqSD}
\end{equation}
where $J$ is the determinant of the deformation gradient tensor, $G$ is the shear modulus of the material  and ${\bf B}'$ is the deviatoric part of ${\bf B}$, ${\bf B}'={\bf B}-{1\over3}{\rm tr}{\bf B}$. The second term, $-p{\bf I}$, is the hydrostatic pressure term

This solid stress model describes a pure elastic effect and does not include losses when the solid moves and deforms.
To allow energy losses a viscous damping term is added. This term is of the same form as used for the fluid stresses, ${{\bm \sigma}'_{s,d}}=2\mu_s {\bf D}$, where $\mu_s$ is a damping coefficient and given the maximum value of the viscosity indicated in Part I. The deviatoric tensor for the solid stress model, including both the elastic and viscous term is:
\begin{equation}
{{\bm \sigma}'_s}={\bm \sigma}'_{s,e}+{\bm \sigma}'_{s,d}.
\label{eqSDt}
\end{equation}

\subsection{Modeling the Nozzle}

The nozzle is modeled as a cylinder, with one end closed and the other end open, represented computationally by an immersed boundary. The melted polymer is injected inside the cylinder by volume and heat sources. This is similar to the idealized injection used in Part I, but here the addition of a rigid immersed boundary (the cylinder) guides the flow in a specific direction. The nozzle has a finite wall thickness and at the nozzle walls the relative velocity is zero. Except for the total amount of polymer injected and the no-slip walls, the velocity profile at the nozzle exit is not specified explicitly. The volume and the heat sources are placed at the top of the nozzle, occupying a cylindrical region smaller than the inside diameter of the nozzle. The immersed boundary forming the nozzle, as well as the volume and the heat source, move with a prescribed velocity.

To identify the solid region forming the nozzle walls, a new index function, $I^{nozzle}$, is defined, where 1 denotes the cells occupied by the solid and 0 everything else. In addition to specifying the velocity inside the solid region (zero for a stationary nozzle and equal to its velocity if it is moving),  the pressure equation is modified slightly to set the normal velocity at the nozzle wall to the correct value. To ensure that the normal pressure gradient at the wall is approximately zero, the pressure equations for the cells occupied by the solid body are replaced in the following way: 
\begin{equation}
{p_{i,j,k}}=\left\{
\begin{array}{rcl}
{{\Sigma(1- I_{l,m,n}^{nozzle}) p_{l,m,n}} \over {\Sigma I_{l,m,n}^{nozzle} }}, & {\Sigma(1- I_{l,m,n}^{nozzle})} \neq 0;\\
\\
{1\over 6}\Sigma p_{l,m,n}, &  {\Sigma(1- I_{l,m,n}^{nozzle})}  = 0,
\end{array}
\right.
\end{equation}
where the subscript $(l,m, n)$ traverses the six cells surrounding $(i,j,k)$. Thus, the pressure in cells just inside the nozzle is set to the average value of the pressures of the neighboring cells outside the nozzle, and the pressure of cells fully inside the nozzle is set to the average of the surrounding cells. The velocity in the cells inside the solid body is set to the nozzle velocity, ${\bf u}={\bf u}_{nozzle}$. 
The ``height'' of the nozzle is taken to be 1.5 times of its diameter, which is enough to let the flow inside the nozzle develop fully. 
The locations of the nozzle walls and the sources are moved at the begin of each time step.

\begin{table}[H]
\caption{Material Properties for the polymer}\label{ParametersTable}
\begin{tabular}{ccc}
\toprule
Parameter								& value						&Ref. or Eq.\\
\midrule
Glass transition point $T_g$ ($^\circ$C)  		& 55						&\citep{farah2016}\\
Solidus(melting) point $T_m$ ($^\circ$C)  	& 165						&\citep{farah2016}\\
Density $\rho_a$ ($kg/m^3$)				& 1170$\sim$1269 (function) 		&\citep{grassia2011}\\
Young's modulus $E$ ($Pa$)				& $3.500\times10^9$			&\citep{farah2016}\\
Poisson's ratio $\nu$ (-)					& 0.36						&\citep{farah2016}\\
Shear modulus $G$ ($Pa$)				& $1.287\times10^9$			&$E/2(1+\nu)$\\
Bulk modulus $\kappa$ ($Pa$)				& $4.167\times10^9$			&$E/3(1-2\nu)$\\
\bottomrule
\end{tabular}
\end{table}

\begin{figure}[H]
\centerline{\scalebox{0.35}{\includegraphics{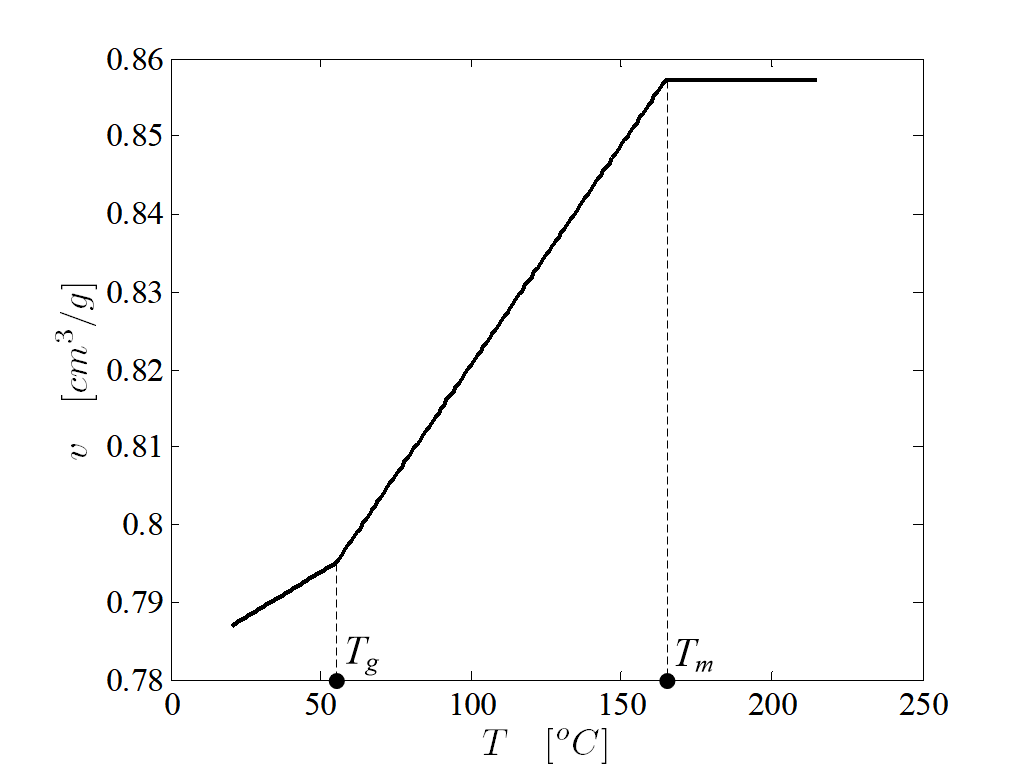} }}  
\caption{The dependency of the specific volume of the polymer on the temperature.}
\label{Figdensity}
\end{figure}

\subsection{Improved Numerical Algorithm}
The momentum equations are solved using the implicit projection scheme introduced in Part I.
The projection scheme leads to a non-separable Poisson equation for a pressure-like potential, $\tilde p$, that enforces incompressibility.
The potential differs slightly from the pressure, since the viscous term is treated implicitly in the predictor step rather than the correction step (\cite{kim1985}). The pressure and the potential are related by $p-{\widetilde p}=[\Delta t/(2Re)]\nabla ^2{\widetilde p}$ where it is clear that the difference increases  with $\nabla ^2{\widetilde p}$. For the new method  the pressure gradient is very high inside the nozzle due to the relative high viscosity and velocity, and to deal with that, the projection scheme is modified slightly, but  the implicit scheme is unchanged. The momentum equation is thus split as:
\begin{equation}
{\rho^{n+1}{\bf u}^*-{\rho^n{\bf u}^n}\over {\Delta t}}=
{1\over 2}{\bf T}^*+{1\over 2}{\bf T}^n+\nabla \cdot I_s {\bm \sigma}'^{n}_{s,e}- \nabla {\widetilde p}^n - {\bf A}^n + {\bf S}^n,
\label{PJ1}
\end{equation}
and
\begin{equation}
{\rho^{n+1}{\bf u}^{n+1}-{\rho^{n+1}{\bf u}^*}\over {\Delta t}}=-\nabla {\delta {\widetilde p}}.
\label{PJ2}
\end{equation}
where the stress term ${\bf T}=\nabla \cdot [{\bm \sigma}'_f+({\bm \sigma}'_{s,d}-{\bm \sigma}'_f) I_s]$.
Here, the pressure gradient at time $n$ is kept in the first split equation, and then the original pressure in the Poisson equation is replaced by ${\delta {\widetilde p}}$. Finally, the updated pressure is ${\widetilde p}^{n+1}={\widetilde p}^n+\delta {\widetilde p}$, and the updated velocity is ${\bf u}^{n+1}={\bf u}^*-({\Delta t /\rho^{n+1}})\nabla\delta {\widetilde p}$.

Time integration is done using a third order Runge-Kutta method and spatial derivatives are approximated using second order centered finite difference approximations.
The pressure Poisson equation is solved using a  Bi-CGSTAB method and a TFQMR (Transpose-Free Quasi-Minimal Residual) method is used to solve the split momentum equation. 

\section{Results}
The material simulated is PLA, as in Part I. The thermal and fluid related material properties can be found in Part I. Here only  the parameters used in the modeling of the solid are listed, see Table \ref{ParametersTable}. The relationship between the specific volume $v$ and the temperature $T$ for PLA is taken from \citep{grassia2011}, except that the specific volume is set to a constant when the temperature is higher than the melting point $T_m$. Thus, $v(T>T_m)=v(T_m)$, as shown in Figure \ref{Figdensity}.

In addition to simulations using the correct value of the Young's modulus, a few cases using lower values have also been examined. Those are easier to compute and allow the mathematical model and the numerical algorithm to be checked at lower computational cost.

\subsection{Injection with an Improved Nozzle Model}
To test the new model for the nozzle,  a short filament is laid down on a flat surface. A nozzle with an internal diameter of $0.4mm$, moving along a straight will deposit  a cylindrical filament with a nominal diameter of $D=0.78mm$, for the material feed rate and nozzle speed given in Part I. The computational domain is a hexagonal box with dimensions  $3.13 \times 1.76 \times 1.17 mm$. The exit of the nozzle was placed a distance $D$ above the bed. The temperature of the material injected and the body of the nozzle are set to a  fixed injection temperature of $215^\circ$C, and the bed is at $40^\circ$C.

\begin{figure}
\centerline{\scalebox{0.35}{\includegraphics{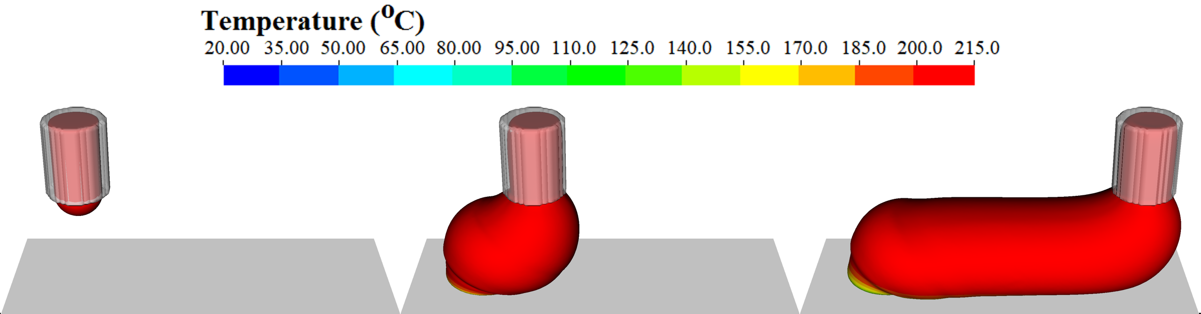} }}  
\caption{Three frames from a simulation of one filament deposited using the new nozzle model. The color shows the temperature on the surface, and the semitransparent gray volume shows the nozzle body.}
\label{NozzleFig}
\end{figure}

Three frames from a simulation of the formation of a short filament by the extrusion of a hot polymer, using the new nozzle model, are shown in Figure \ref{NozzleFig}. In each frame, the immersed boundary forming the nozzle is shown by a semitransparent gray volume and the hot polymer is initially red. In the first frame the nozzle is stationary and the polymer consists of a short cylinder with a hemispherical end that sticks out from the nozzle. In the second frame the polymer has hit the bed and the nozzle has just started to move, and in the third frame a short filament has been formed. The filament cools down slightly at the point where it initially contacted the bed, but for most part the time is too short for any significant cooling to have taken place. Since material (and heat) is constantly added, the moving nozzle is always full.

The main difference between the nozzle model used here and the simple source used in Part I, is that the polymer is injected in a specific direction (downward) through a given cross section, and that heat is added by keeping the nozzle body at a constant temperature. Compared to the results in Part I, here the cross section of the filament is more elliptical, and the front of the newly deposited filament extends ahead of the moving nozzle. The shape of the nozzle and the injection rate are the only conditions that are imposed, and the flow and the geometry of the extruded filament are fully determined by the shape and motion of the nozzle and found as part of the solution.

\begin{figure}
\centerline{\scalebox{0.35}{\includegraphics{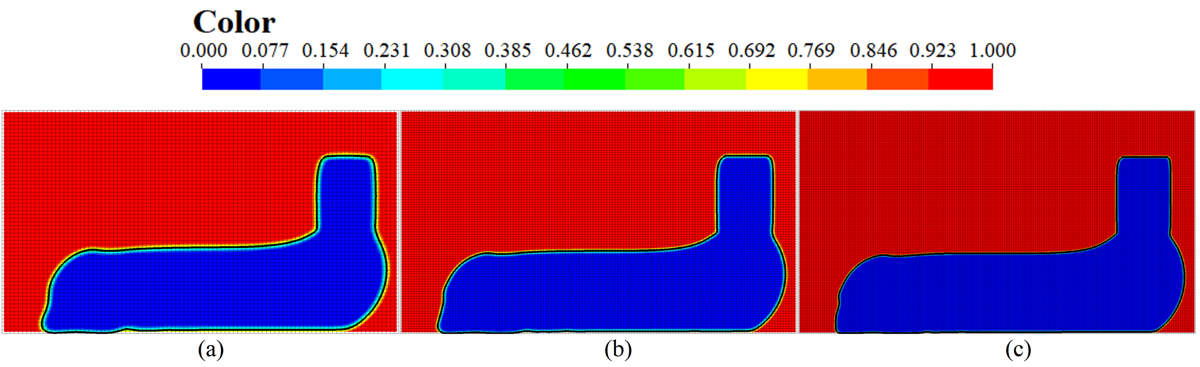} }}  
\caption{A cut through the  longitudinal section of the finished filament for three different resolutions. The color shows the polymer and the air respectively, and the black line is the interface. The resolution of the full computational domain is: (a) $96 \times 60 \times 36$, (b) $144 \times 90 \times 54$ and (c) $216 \times 135 \times81$, grid points.}
\label{NozzleConvergence0}
\end{figure}

To examine the behavior of the nozzle model further, a grid-refinement study, was carried out, using three levels of grid resolutions. The coarsest grid has  $96\times60\times36$ grid points, which is 1.5 times lower than used in Figure \ref{NozzleFig}, and the finest grid has $216\times135\times81$ grid points, which is 1.5 times higher. Figure \ref{NozzleConvergence0} shows the profiles in the middle longitudinal section, all at the same time. The filaments, as indicated by the blue color, are very similar, and key features, such as the start shape on the left and the shape under the moving nozzle, suggest convergence under grid refinement.

\begin{figure}
\centerline{\includegraphics[scale=0.4]{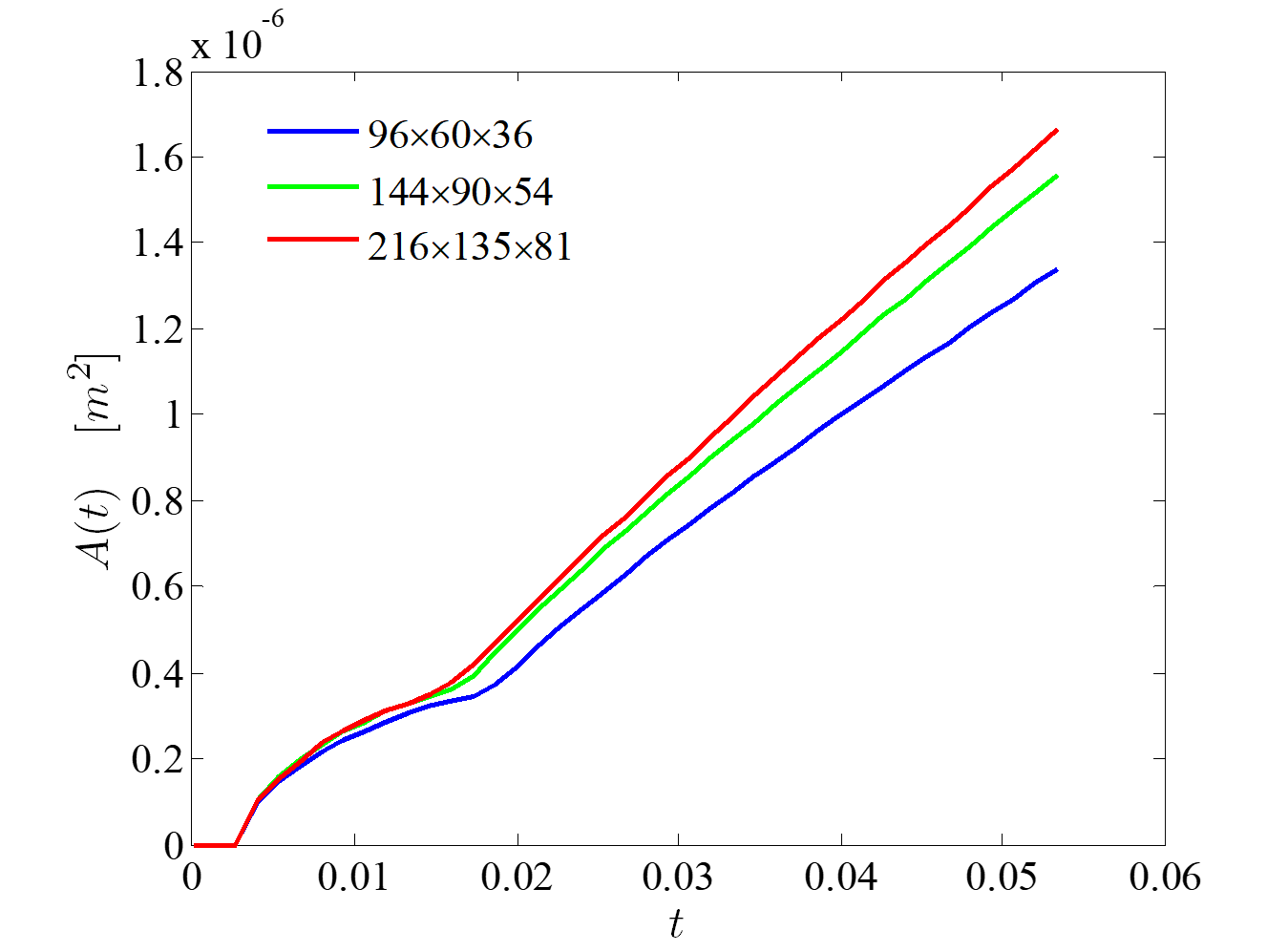}} 
\caption{Contact area $A(t)$ between the filament and the bed versus time for three different resolution, $96\times60\times36$, $144\times90\times54$ and $216\times135\times81$ grid points.}
\label{NozzleConvergence1}
\end{figure}

To examine the convergence more quantitatively, the contact area $A(t)$ between the filament and the bed is plotted versus time in Figure \ref{NozzleConvergence1} for the three resolutions. The contact area increases almost linearly with time, except at the start, where the polymer first makes contact with the bed. The $A(t)$ curves for the finer two resolutions are close, suggesting a converging solution. The total volume $V(t)$ of the injected filament body (not shown) has also been computed and found to be essentially the same for the different resolutions.

\subsection{Solidification, Volume Change and Stresses in the Solid}

To examine the performance of the numerical method when the solidification, shrinkage and  stresses in the solid are included, the configuration introduced in Part I is used. The computational domain is $3.13\times2.35\times1.57 mm$, and two short filaments are laid down one on top of the other. Since the injection time is very short so no significant cooling takes place during the injection, the computations presented here start after the filaments have been formed. Because realistic values for the Young's modulus require very small time steps, a value that is 100 times smaller is used for most of the studies in this section. Thus $\Delta t=6\times10^{-6}-10^{-5}s$ here, but ten times smaller for the simulations in section 3.3. As shown later, higher values for the Young's modulus lead to qualitatively similar evolution (but higher stresses). The focus is on quantities that depend on the changes in the temperature $T$, namely the Jacobian $J$, the total volume $V$,  the volume shrinkage ratio ${\Delta V / {V_{ref}}}$, and the mean stress $\sigma_p$. Of these quantities, the Jacobian directly measure the shrinkage and is therefore of major significance. Since the volume change is only due to the change of density with temperature, the Jacobian also only depends on the temperature, $J=J(T)$.

\begin{figure}
\centerline{\scalebox{0.30}{\includegraphics{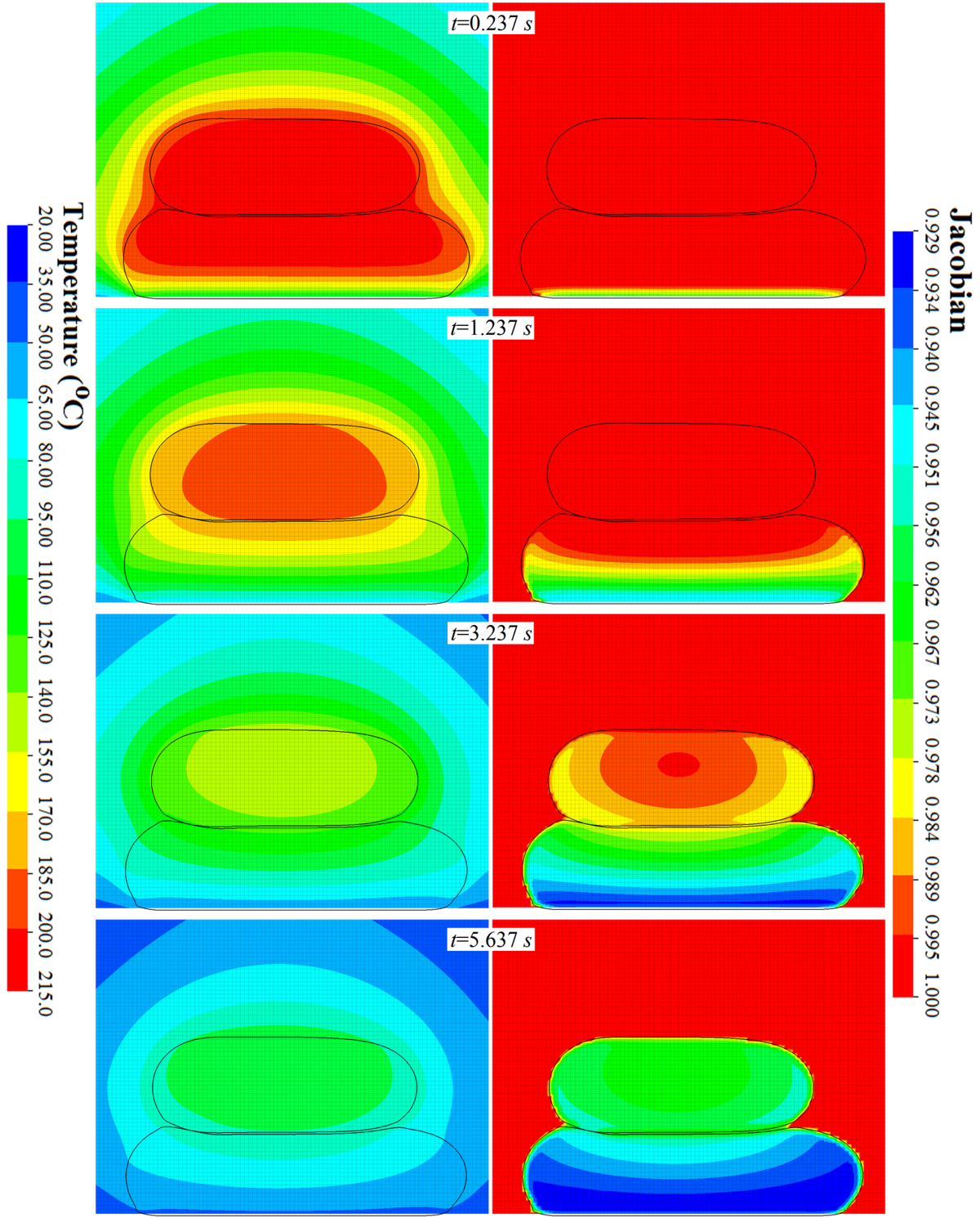} }}
\caption{The evolution of the temperature and Jacobian with time in a simulation with Young's modulus equal to $3.5\times10^7 Pa$. The frames in the left column show the temperature in a longitudinal section through the middle of the domain at different time, and the frames in the right column show the Jacobian in the same plane at the same times.}
\label{TJ}
\end{figure}

Figure \ref{TJ} presents four frames showing the temperature and the Jacobian in a plane cutting through the middle longitudinal section, at four times, from a simulation using a Young's modulus equal to $3.5\times10^7 Pa$. Below the melting point the Jacobian is a function of the temperature only, so once it is less than unity solidification is taking place.  The first frame shows that both filaments are still at nearly the injection temperature, except for a very small region near to the bottom where the polymer has solidified and the Jacobian is less than $1$.  
After 1 second, the bottom filament and part of the top filament have cooled a little bit and the Jacobian shows a larger solidified region in the bottom filament.  No part of the top filament has started to solidify, since its temperature is still higher than the melting point $T_m$. After $3$ seconds both filaments have cooled down significantly and the Jacobian shows that a significant part of the top filament is solid. After 5 seconds (bottom frames) both filaments have become solid and continue to cool down, leading to further shrinkage, particularly of the bottom filament. As the figure shows, the solidification, as indicated by the Jacobian, progresses from the bottom of the filament and from the edges to the center, directly following the cooling of the filament.

\begin{figure}
\raggedleft
\subfloat[]{ 
\label{fig:VJ1} 
\begin{minipage}[c]{0.5\textwidth} 
\centerline{\includegraphics[scale=0.35]{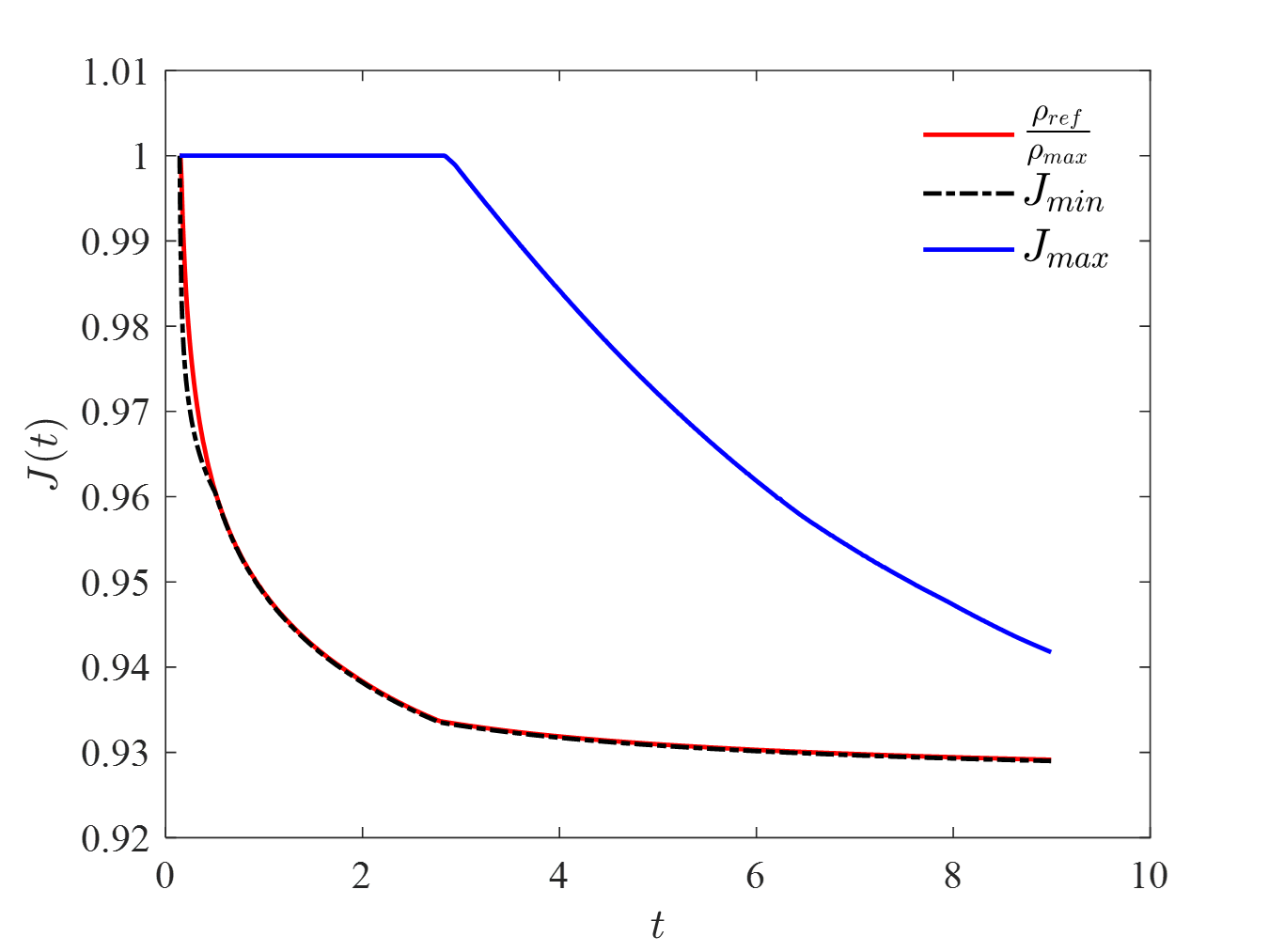}} 
\end{minipage}
}
\raggedright
\subfloat[]{ 
\label{fig:VJ2} 
\begin{minipage}[c]{0.5\textwidth} 
\centering{\includegraphics[scale=0.35]{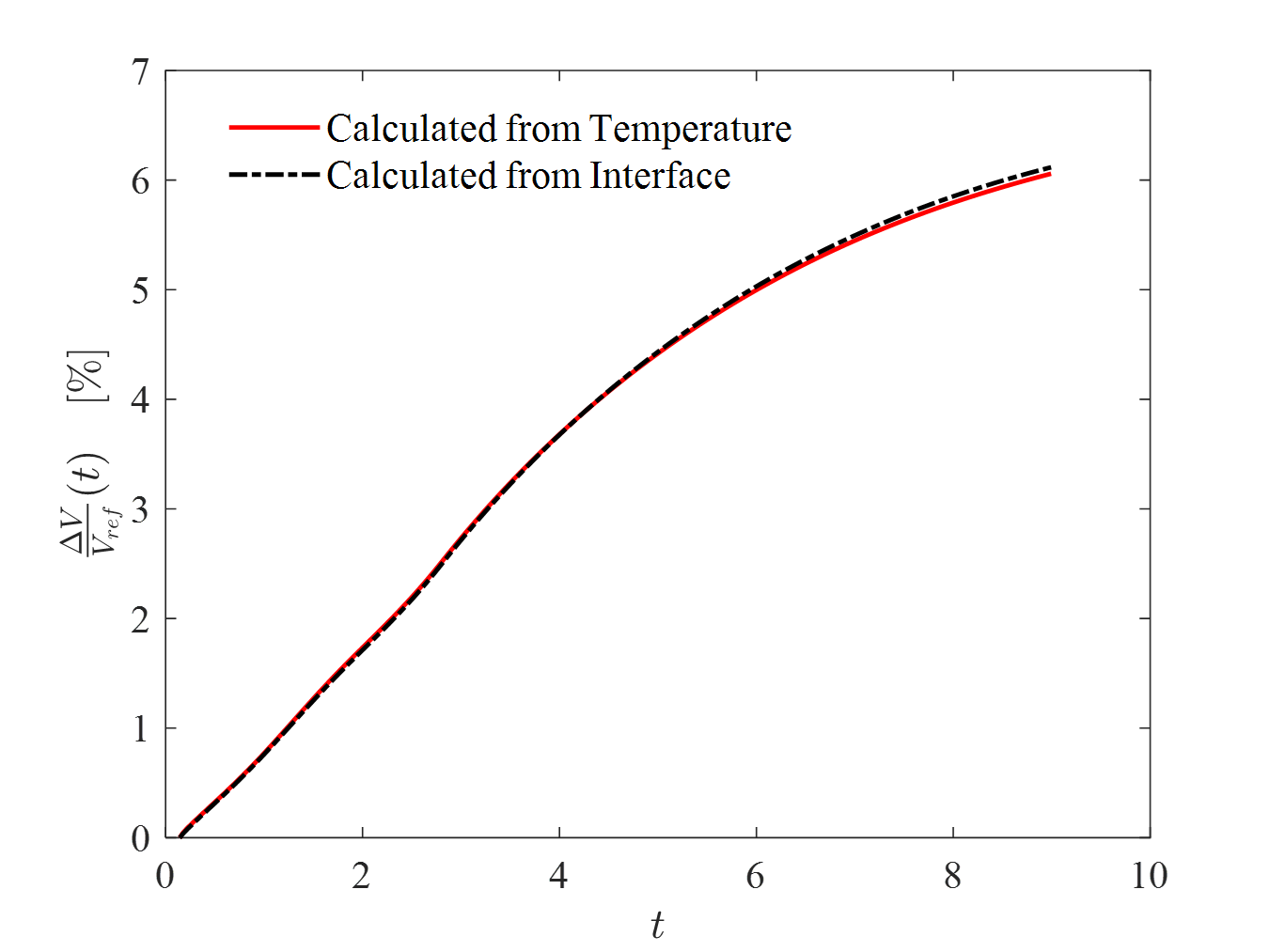}} 
\end{minipage} 
}
\caption{(a) $J_{min}$ and $J_{max}$ in the polymer versus time, as well as the reciprocal density change ratio.  (b) Volume shrinkage ratio, ${\Delta V \over {V_{ref}}}(t)$ versus time, calculated by equations (\ref{VolumeEq}).}
\label{VJ}
\end{figure}

The evolution of the Jacobian for the polymer is examined in a more quantitative way by plotting the minimum and maximum value versus time. The Jacobian can be computed in two ways, directly from the deformation gradients ($J=det({\bf F})$), or by using that the density is directly related to the Jacobian by $J \rho=\rho_{inj}$, or $J= \rho_{inj} / \rho$, where $\rho_{inj}$ is the injection density, which is equal to the density $\rho_{ref}$ at solidification, as shown in  Figure \ref{Figdensity}. 
In  Figure \ref{VJ}(a) the maximum value of the Jacobian $J_{max}$  and the minimum value $J_{min}$ are plotted versus time, computed in two ways as described above. Both give essentially the same results, although a slight difference can be seen at the early time when the solidified volume of the filament is very small. Before the polymer starts to solidify it is not allowed to shrink, so $J_{max}=1$ as long as any part of the filament is still at a temperature higher than the solidification temperature. Thus, $J_{max}$ starts out as unity, until about time 3 $s$, when both filaments have solidified. The minimum value levels out as the temperature of the bottom part approaches that of the bed, and the maximum value would eventually level out also at the same value, once both filaments have reached the bed temperature.

\begin{figure}
\centerline{\scalebox{0.35}{\includegraphics{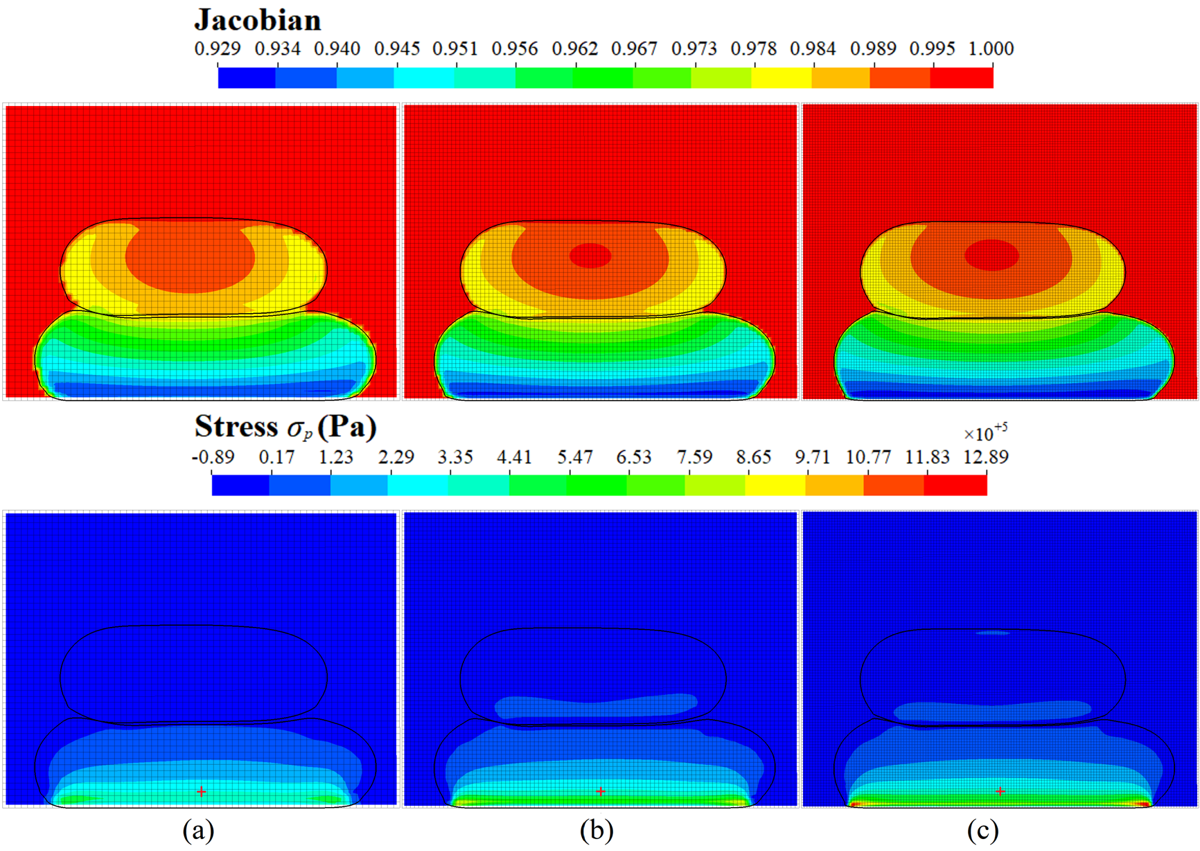} }}  
\caption{The Jacobian and the mean stress at time $t=3.237 s$, in a longitudinal section through the middle of the filaments, for three different resolutions: (a) $64\times48\times32$, (b) $96\times72\times48$ and (c) $144\times108\times72$ grid points.}
\label{JSigm_E7}
\end{figure}

The volume shrinkage ratio ${\Delta V /{V_{ref}}}$, defined as the ratio of the total shrunken volume $\Delta V$ and the total extruded polymer volume $V_{ref}$,
can also be found in two ways, since the shrunken volume can be found in two ways. The rate of shrinkage (which only depends on the temperature) can be integrated over the filament and time or the volume by integrating over the surface.  The change between times $t_s$ and $t$ is therefore: 
\begin{equation}
\Delta V=-\int_{t_s}^t \int_{\Omega} \nabla\cdot {\bf u}dvdt; \quad  \hbox{and} \quad
\Delta V=- \left(\oint_{\partial  \Omega} {1 \over 3}{\bf x}_f \cdot{\bf n}da \Big|_t-\oint_{\partial  \Omega} {1 \over 3}{\bf x}_f \cdot{\bf n}da \Big|_{t_s}\right).
\label{VolumeEq}
 \end{equation}
Here, $\Omega$ and $\partial \Omega$ denotes the filament and the interface separating the polymer and air, respectively.  ${\bf x}_f$ is the coordinates of points on the interface and {\bf n} is a unit normal vector. $t$ is the current time and $t_s$ is the time when the injection has just finished. Obviously, both approaches should give the same results. Figure \ref{VJ}(b) shows the volume change ratio versus time found in those two ways. The slope of the curve for the volume shrinkage ratio, closely related to the shrinkage rate, depends on how rapidly the polymer cools down as well as the slope of the specific volume versus temperature curve in Figure \ref{Figdensity}. The curve indicts that the cooling rate is nearly constant until about $3-4s$. After that the cooling rate slows down and the slope of the shrinkage ratio curve slows down as well.

\begin{figure}
\raggedleft
\subfloat[]{ 
\label{fig:V} 
\begin{minipage}[c]{0.5\textwidth} 
\centerline{\includegraphics[scale=0.30]{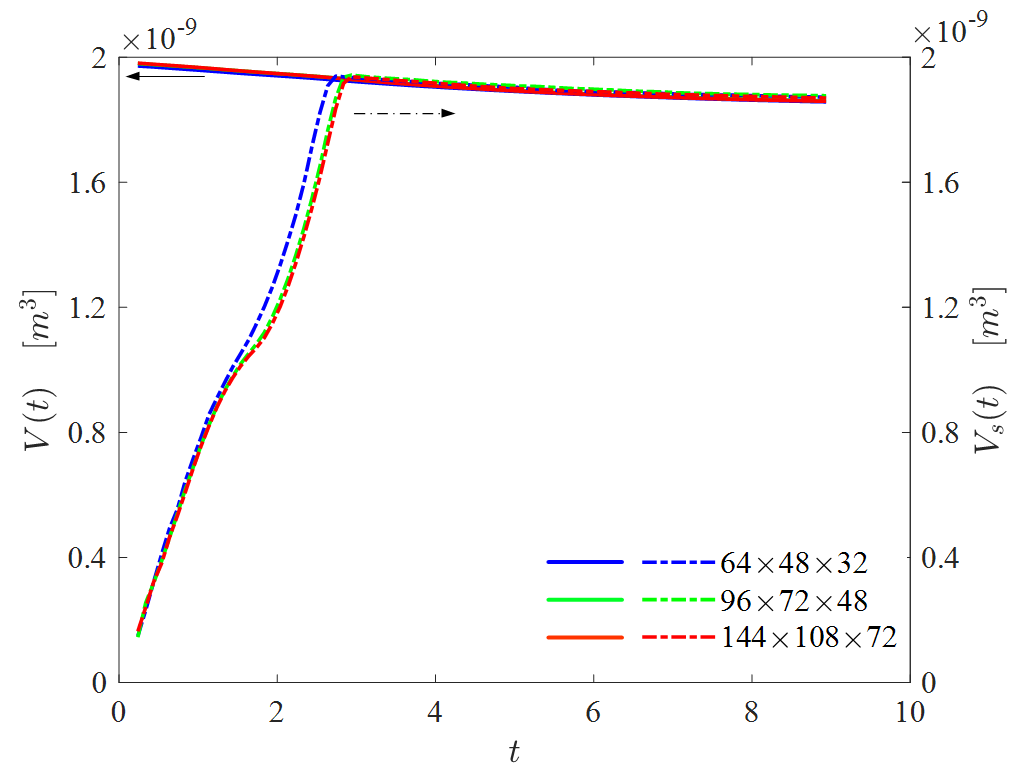}} 
\end{minipage}
}
\raggedright
\subfloat[]{ 
\label{fig:Sigma} 
\begin{minipage}[c]{0.5\textwidth} 
\centering{\includegraphics[scale=0.30]{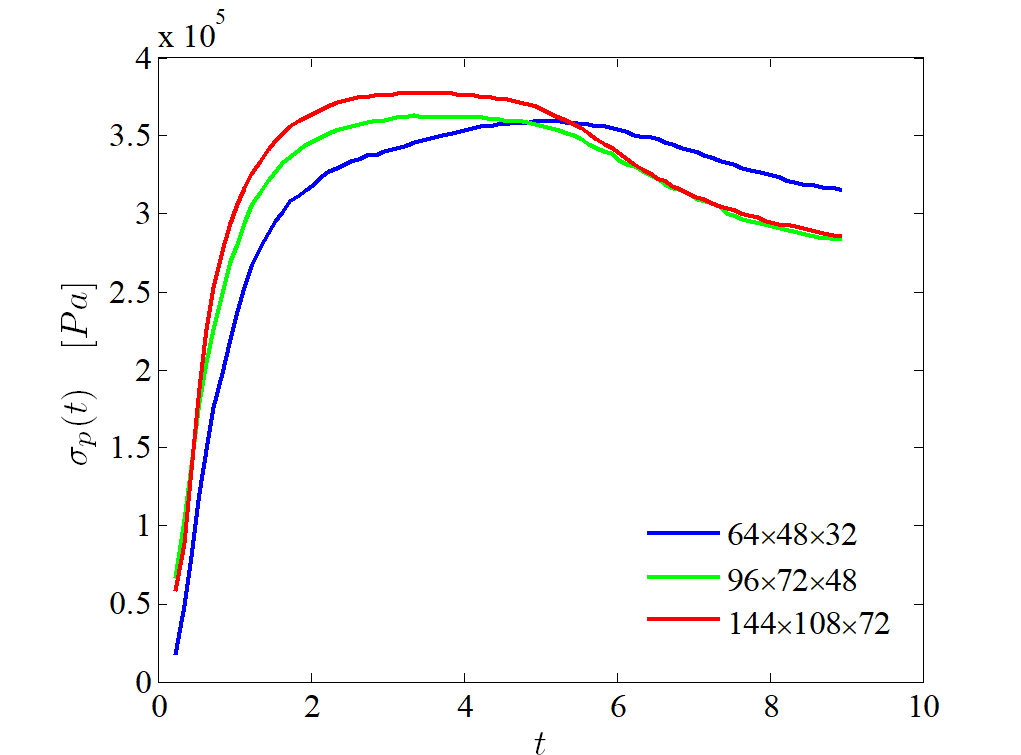}} 
\end{minipage} 
}
\caption{(a) The total volume $V(t)$ (solid lines) and the solidified volume $V_s(t)$ (dashed lines), versus time. (b) The mean stress $\sigma_p(t)$ versus time, at one point inside the body of the bottom filament.}
\label{VSigma}
\end{figure}

The convergence of the solidification model is examined in Figure \ref{JSigm_E7}, where the Jacobian and the mean stress
\begin{equation}
\sigma_p\equiv{\sigma_{xx}+\sigma_{yy}+\sigma_{zz} \over 3},
\label{MeanStress}
\end{equation}
in a longitudinal section through the middle of the computational domain are plotted, at time $t=3.237s$, as computed on three different grids with $64\times48\times32$, $96\times72\times48$ and $144\times108\times72$ grid points. The Jacobian is lowest at the bed and increases toward the middle of the top filament, where the temperature is highest. Overall the agreement is good but the low resolution results in slightly smaller values. The mean stresses exhibit a similar trend, with slight differences seen for the lowest resolution. The comparison is more quantitative in Figure \ref{VSigma}(a), where the  total volume $V(t)$ and the solidified volume $V_s(t)$ are plotted versus time for all three resolutions. The total volume (solid lines) is gradually reduced due to shrinkage as the polymer cools down, and the fraction that is solid (dashed lines) increases until about 3 $s$, when everything is solid and the total volume and the solidified volume are the same. While the total volume is nearly the same for all resolutions, the lowest resolution over-predicts slightly the solidified volume. Figure \ref{VSigma}(b) shows the mean stress at one point, marked by a red cross in Figure \ref{JSigm_E7}. As the filaments cool down the stress first increases rapidly and then gradually relaxes as the temperature becomes uniform. Although there is a slight difference between the two finest resolutions, the results are clearly converging.

\subsection{Real Material Properties}

\begin{figure}
\centerline{\scalebox{0.35}{\includegraphics{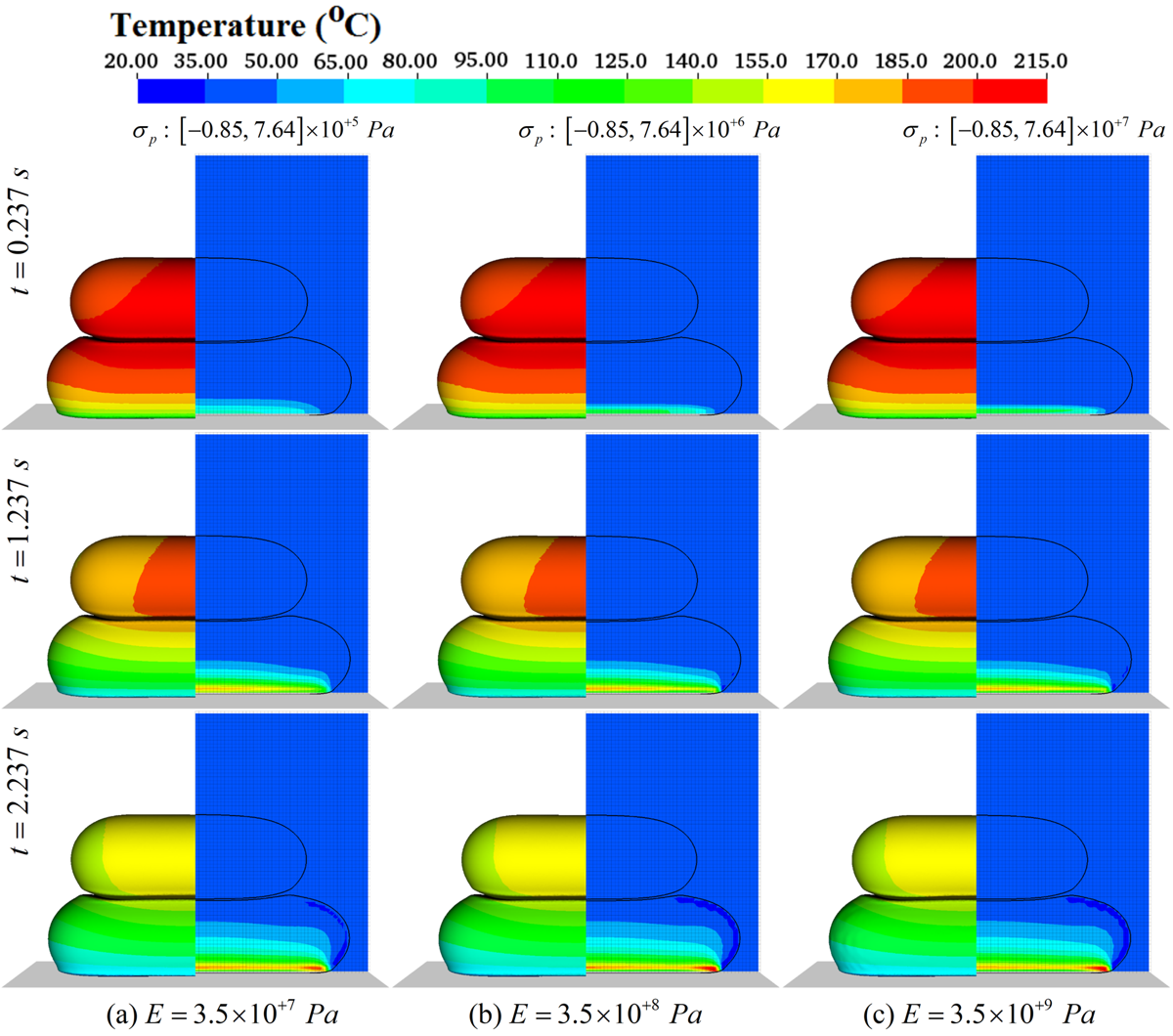}  }}  
\caption{Comparison of the shape, temperature and mean stress for simulations with different Young's modulus: (a) $3.5\times10^7 Pa$, (b) $3.5\times10^8 Pa$, and (c) $3.5\times10^9 Pa$ at times $t=$0.237, 1.237 and 2.237 $s$. For each frame the filaments and the temperature on their surfaces are shown on the left, and the mean stress, $\sigma_p$ in the middle-longitudinal-section  on the right. }
\label{Comp789Front}
\end{figure}

The simulations in the previous section, with a Young's modulus equal to $3.5\times10^7 Pa$, produced results that appear to converge under grid resolution, and look physically correct. However, realistic values for material of practical interest are about 100 times larger and here the method is tested for Young's modulus equal to $3.5\times10^7 Pa$, $3.5\times10^8 Pa$, and $3.5\times10^9 Pa$. The highest value corresponds to PLA feedstock commonly used for FDM/FFF. Figure \ref{Comp789Front} shows the three-dimensional shape and the surface temperature on the left and the mean stress on the right, at three times for the three values of the Young's modulus. The filament shapes and the temperature fields are essentially the same. The stress distribution is also very similar, but the values are very different, and increase by about 10 for each increment in the Young's modulus. Specifically, the maximum stress is about $7.64\times10^5$, $7.64\times10^6$ and $7.64\times10^7 Pa$ at time 2.237 $s$ for Young's modulus equal to $3.5\times10^7$, $3.5\times10^8$ and $3.5\times10^9 Pa$, respectively. These comparisons show that the  characteristics of the results for the lower Young's modulus are very similar to those for the higher one.

\begin{figure}
\raggedleft
\subfloat[]{ 
\label{fig:VJ1} 
\begin{minipage}[c]{0.5\textwidth} 
\centerline{\includegraphics[scale=0.35]{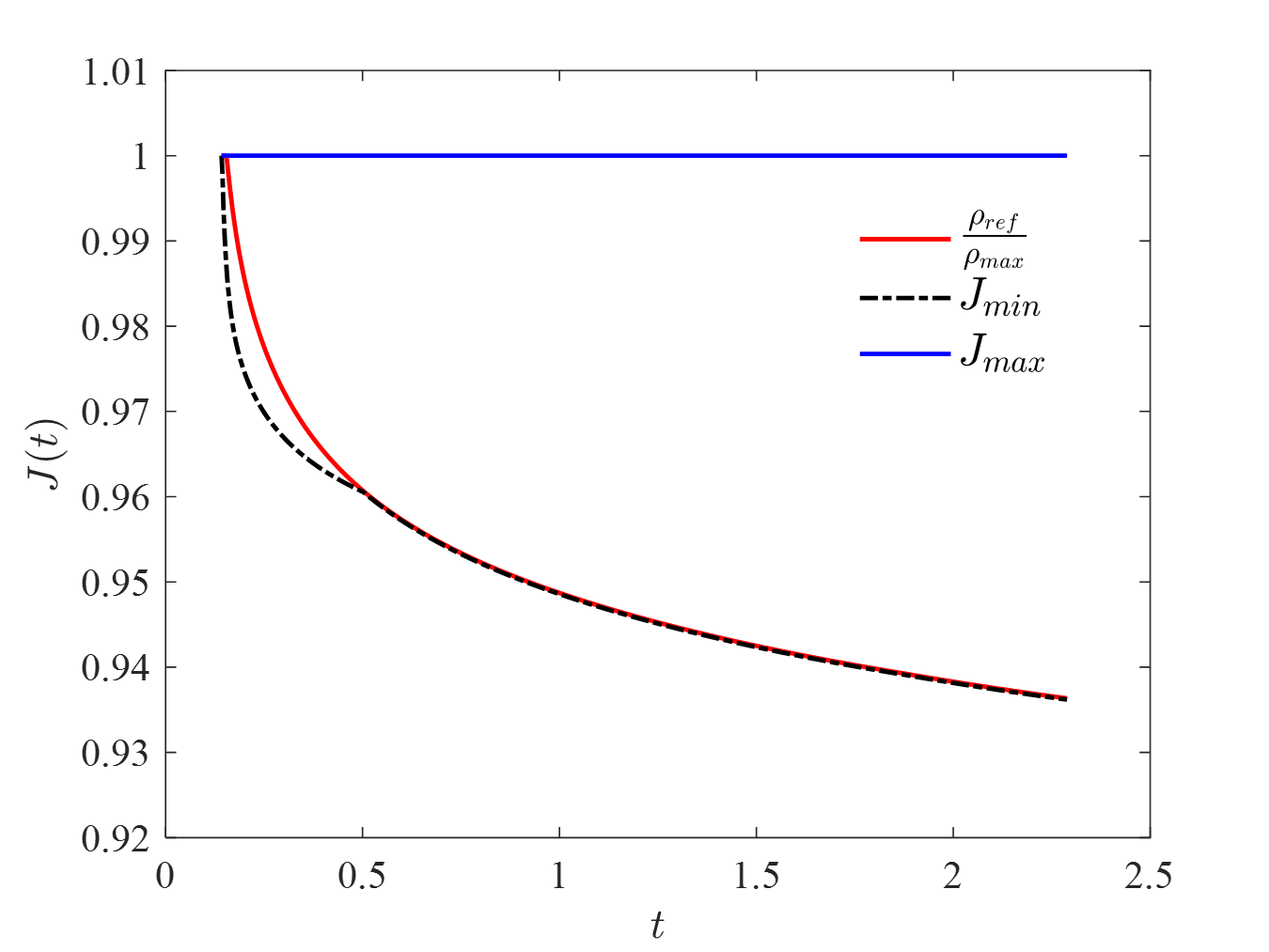}} 
\end{minipage}
}
\raggedright
\subfloat[]{ 
\label{fig:VJ2} 
\begin{minipage}[c]{0.5\textwidth} 
\centering{\includegraphics[scale=0.35]{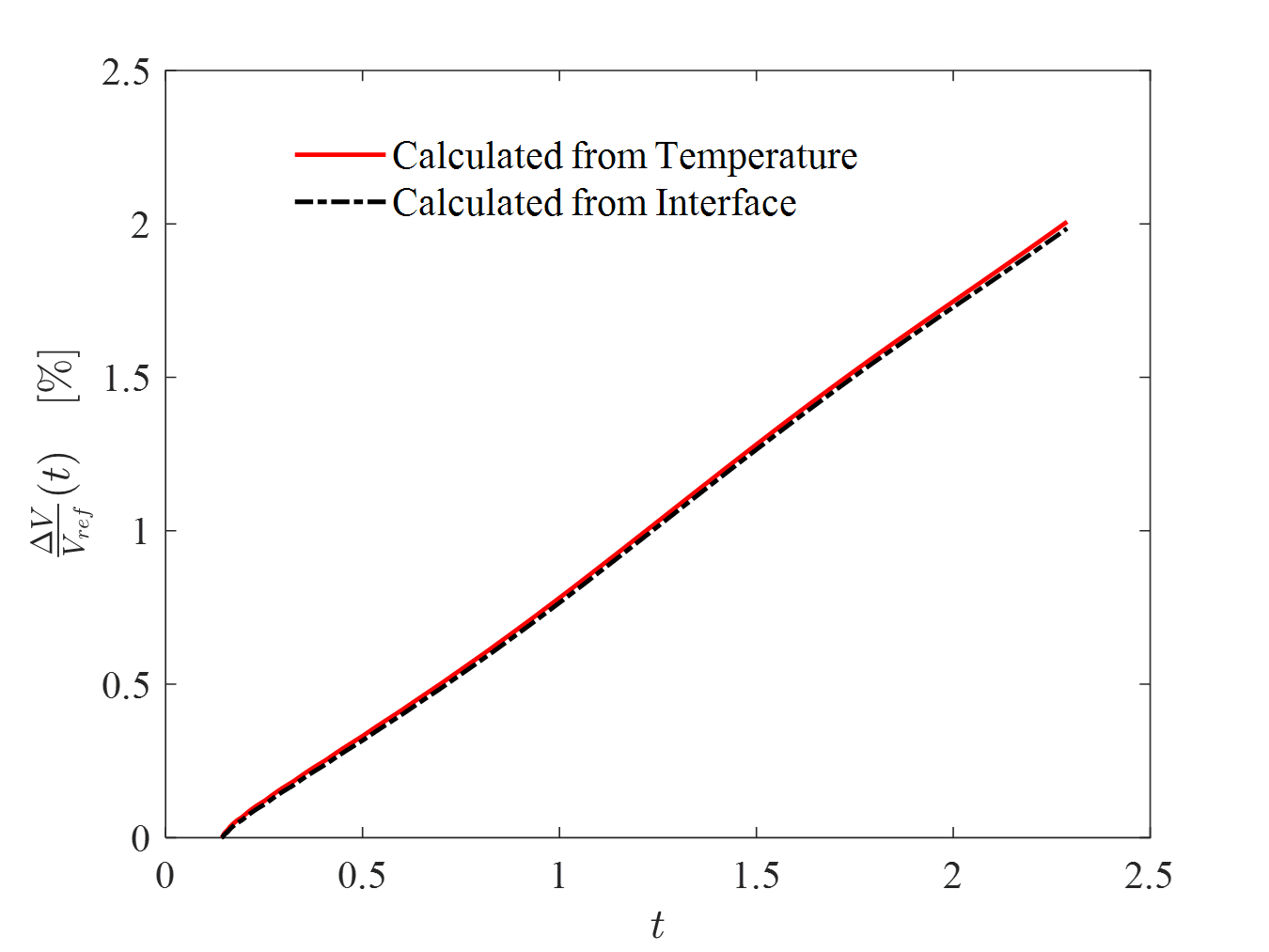}} 
\end{minipage} 
}
\caption{The Jacobian, the density and the volume change rate versus time for material with Young's modulus equal to $3.5\times10^9 Pa$. (a) Comparison between $J_{min}$, ${\rho_{ref} \over {\rho_{max}}}$, and $J_{max}$, (b) Volume shrinkage ratio ${\Delta V \over {V_{ref}}}(t)$ calculated from the temperature field and the interface.}
\label{VJE9}
\end{figure}

To further quantify the performance of the solidification model for the high Young's modulus case, the Jacobian and the volume shrinkage ratio is shown in Figure \ref{VJE9}. Figure \ref{VJE9}(a) shows $J_{max}$, $J_{min}$ and ${\rho_{ref}/{\rho_{max}}}$ versus time.  Here $J_{max}$ stays at unity, showing that parts of the polymer remain unsolidified.  $J_{min}$ is very close to ${\rho_{ref}/{\rho_{max}}}$ after about $0.5s$. For the volume shrinkage ratio, shown in Figure \ref{VJE9}(b),  ${\Delta V / {V_{ref}}}$ calculated by integrating over the interface is very close to the value found from the temperature field. The evolutions of the Jacobian and the volume shrinkage ratio have similar characteristics as the ones for the case with Young's modulus equal to $3.5\times10^7 Pa$, suggesting that the evolution is not fundamentally different and the convergence properties are likely to be the same as for the lower values.

\begin{figure}
\centerline{\scalebox{0.35}{\includegraphics{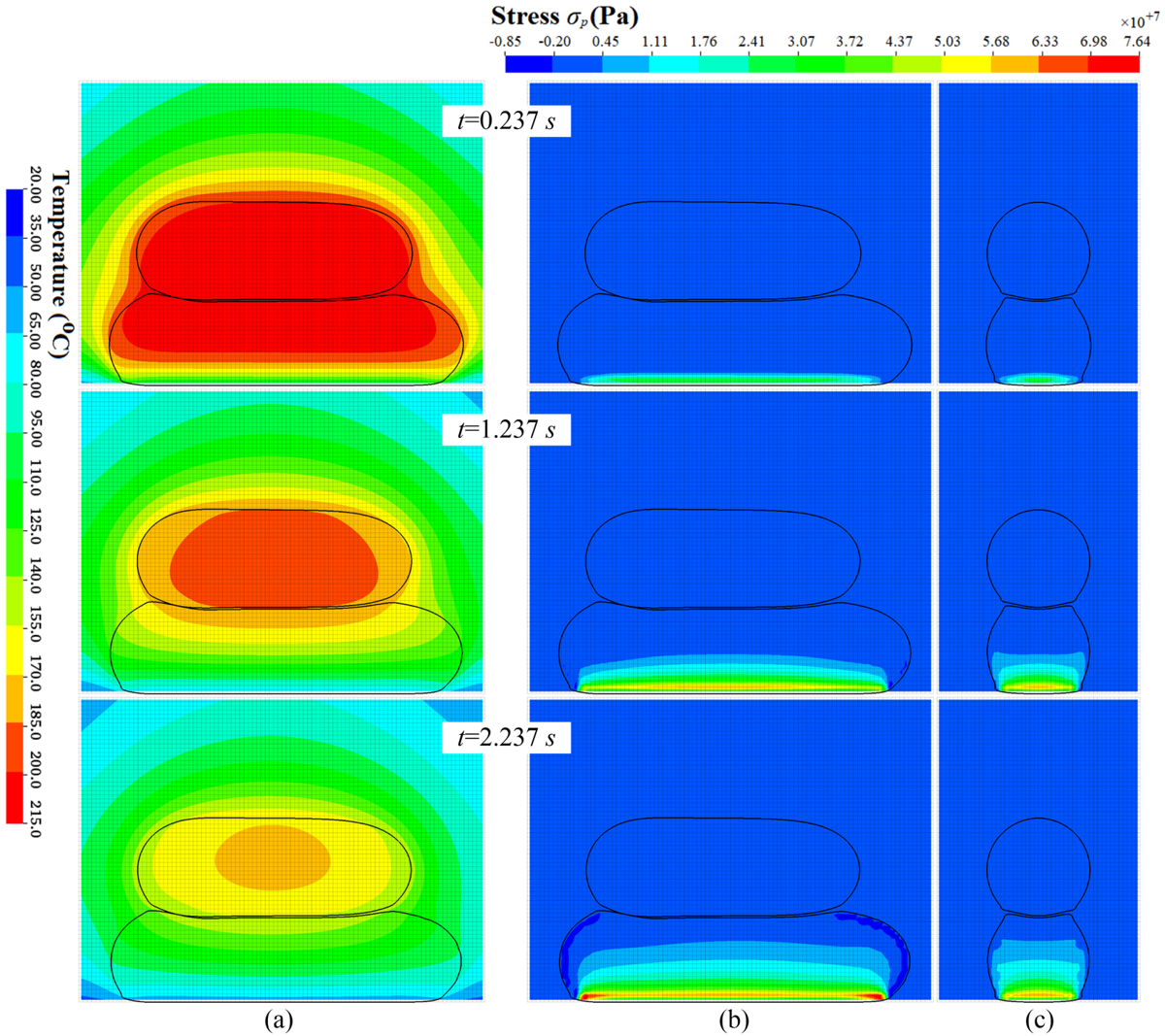}  }}  
\caption{The temperature and the mean stress $\sigma_p$  at times, $t=$0.237, 1.237 and 2.237 $s$, for a material with a Young's modulus of $3.5\times10^9 Pa$. (a) Temperature in the middle longitudinal section, (b) the mean stress in the middle longitudinal section and (c) the mean stress in the middle cross section.}
\label{StressE9}
\end{figure}

Figure \ref{StressE9} shows the temperature at three times in a longitudinal cross section through the middle of the domain, the mean stress in the same plane, and the mean stress in a perpendicular plane cutting through the middle of the filaments.  The temperature field is similar to  the case with Young's modulus equal to $3.5\times10^7 Pa$. At the earliest time the temperature within about $0.1mm$ from the bottom drops to below the melting point $T_m$, showing that the solidification starts from the bottom. The mean stress is high in the low-temperature region, and the maximum stress is about $2.7\times10^7 Pa$. As the cooling progresses, the low-temperature region extends upwards inside the bottom filament and inwards from the edge for the top filament, although it has not solidified yet. The stress is highest at the bottom where the filament has solidified and the temperature is lowest. The maximum stress,  about $6.0\times10^7 Pa$, is seen near the end of the filament.

For the two-filament configuration the bottom filament sticks to the wall, but the rest of the configuration can shrink freely as it cools down. The no-slip boundary condition is applied at the wall so when the filament cools down after solidification the shrinkage is constrained in how it can deform, resulting in residual stresses. Large temperature gradients, where  the shrinkage is different in adjacent parts of the polymer, are also likely to contribute to residual stresses. The high stresses near the end of the filaments are  due to the large linear shrinkage in the longitudinal direction and the high temperature gradient at the ends, due to contact with both the cold bed and the air. For the simulations presented here the bed temperature has been taken to be constant and no attempt has been made to examine the effect of it on the residual stresses. That should, however, be possible and interesting.

\subsection{More Complex Shapes}

\begin{figure}
\centerline{\scalebox{0.35}{\includegraphics{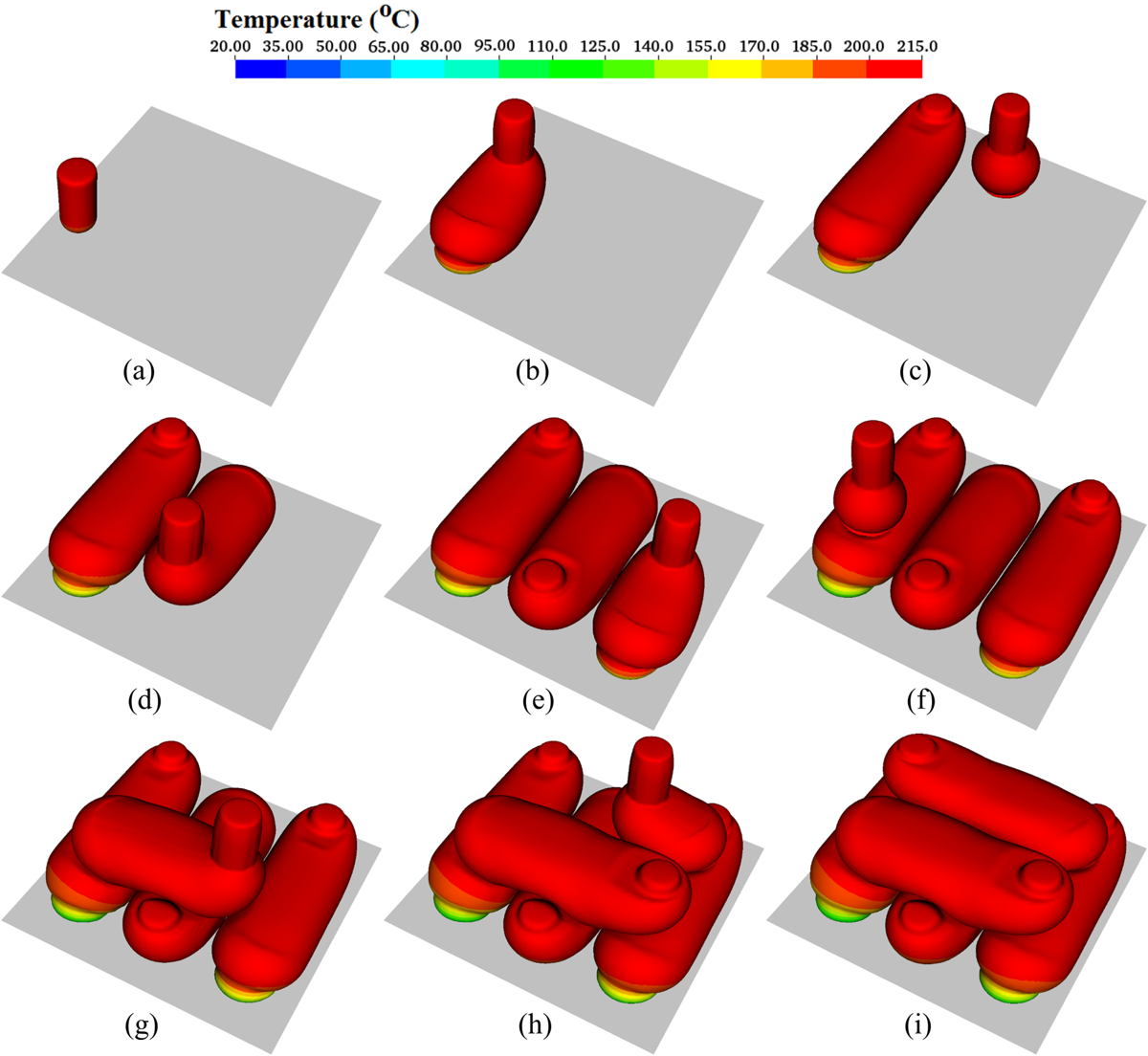}  }}  
\caption{A few frames from a simulation of the fabrication of a two-layer object, with the filaments on top laid down perpendicular to those on the bottom. The computations are done using both the new nozzle and the solidification models. }
\label{LargeCase}
\end{figure}

To demonstrate the use of the method to simulate a section of a full sized object, Figure \ref{LargeCase} shows a few frames from a simulation of the construction of a two-layer object, with the filaments in each layer deposited perpendicular to each other. In this simulation, the new nozzle model is used for the deposition and the solidification model is activated once the polymer has been laid down. The cooling is sufficiently slow so no solidification takes place during the deposition stage and including the solidification model would have had only minor effect on the results, but significantly increased the computational time. To reduce the computational time a Young's modulus of $3.5\times10^7 Pa$ is also used, since the overall results (although not the actual stress values) are likely to be similar to results using a higher value (as seen in figure \ref{Comp789Front}).

The setup is similar to the two-filament cases discussed above, except that a larger computational domain of dimensions $3.13\times 2.35\times 3.13 mm$ is used, resolved by $96\times72\times96$ grid points, to allow  three filaments to be placed at the bottom and two at the top. The injection nozzle is also placed closer to bottom, injecting the bottom and top filament at $0.9D$ and $1.5D$ above the bed, respectively. Once each filament has been laid down, the nozzle and the material inside it are moved to the next injection point, leaving a small ``tap'' at the top of the filament. The first five frames show the injection of the three filament forming the bottom layer, the sixth frame shows the beginning of the injection of the first filament on the top,  the next two frames show the formation of the top layer, and the last frame shows the completed object. While some cooling is visible where the bottom filaments touch the cold bed, most of the filaments are still at nearly the injection temperature. The small taps left on the top of the filaments when the nozzle is moved to the next injection points are likely to decay in time, due to surface tension. However, the injection time (0.3 $s$) is much shorter than the natural decay time for filament of the size and material properties used here.
\footnote{A time scale formed using the filament diameter, the surface tension and viscosity, $\tau=\mu D / \sigma$, gives a decay time scale of $40-100s$, much longer than the time simulated here. }

\begin{figure}
\centerline{\scalebox{0.35}{\includegraphics{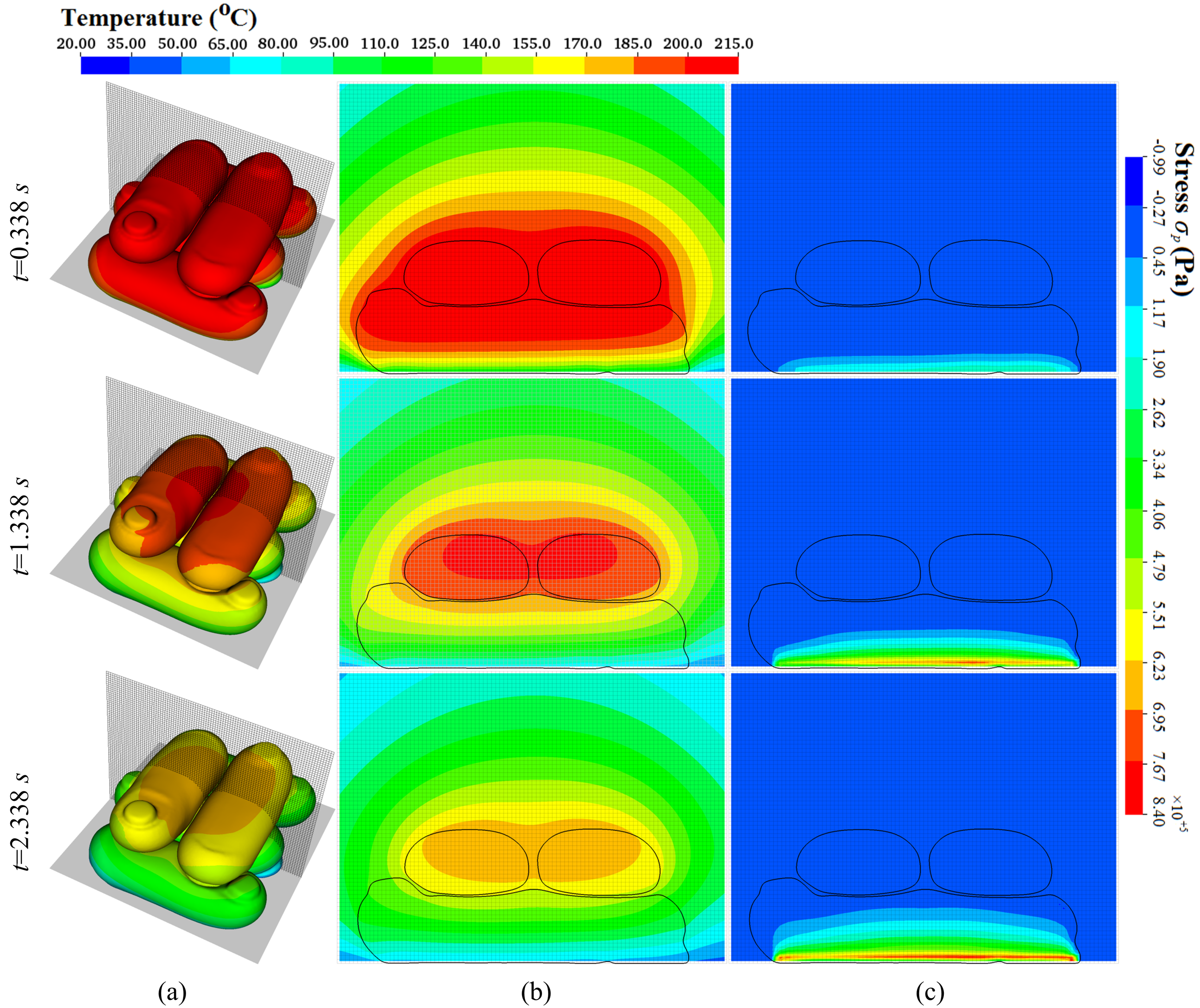}  }}  
\caption{The temperature and the mean stress $\sigma_p$  at times $t=$0.338, 1.338 and 2.338 $s$, for a material with a Young's modulus equal to $3.5\times10^7 Pa$. (a) The shape of the object and the interface temperature, (b) the temperature, and (c) the mean stress in the cross section shown in frame (a).}
\label{LargeCase+TS}
\end{figure}

To examine the solidification process on a longer timescale, the object was cooled down for another 2 $s$ after the injection was completed. Figure \ref{LargeCase+TS} shows three frames at $t=$0.338, 1.338 and 2.338 $s$, respectively, where Figure \ref{LargeCase+TS}(a) shows the shape and surface temperature, and Figures  \ref{LargeCase+TS}(b) and (c) shows the temperature and mean stress $\sigma_p$, respectively, in the cross section shown by a gray plane in frame (a). The main aspects of the cooling and the solidification are similar to what was seen for the two-filament cases, except that there is a small fold near the starting end of each filament which slightly affects the stress close to the bed. 

\section{Conclusions}
A computational framework for FDM/FFF for the flow of a molten  polymer through a nozzle, its deposition and solidification, and the resulting stresses due to the shrinkage as the solid cools down is presented. Grid refinement studies show that converged results can be obtained for conditions encountered in real FDM/FFF processes. While the material models used here capture the expected behavior, more complex models may be necessary to achieve complete agreement with experiments. It is likely, for example, that the reorientation and disentanglement of the polymers as the flow leaves the nozzle plays a significant role in the strength of the bond between adjacent fibers (\cite{McIlroyOlmsted2017, McIlroyOlmsted2017-2}). This can be modeled by solving equations for the evolution of the conformation tensor and modifying the stress tensor to account for its effect, as has been done for simulations of viscoelastic flows, by \cite{Nooranidoostetal2016} and \cite{IzbassarovMuradoglu2015}, using an approach similar to ours.  For low Weissenberg numbers they used the semi-analytical method developed by \cite{SarkarSchowalter2000}, and for higher values they evolved the logarithm of the conformation tensor (\cite{FattalKupferman2005}). Other extensions include adding models for the interdiffusion of the polymer melt, which is also likely to be possible.

The approach described here, and extensions to improve the agreement of the results with experimental observations, is always going to be computationally demanding. Thus, it is unlikely that the method can be used for routine simulations of the fabrication of a full size artifact. It is, however, already capable of describing the fabrication of a relatively simple objects (see part I, \cite {Xiaetal:RPJ:2017} and figures \ref{LargeCase} and \ref{LargeCase+TS} here), and further refinements of the method and faster computers will increase the object size. Thus, it is envisioned that the method will be used to generate samples that can be tested and compared to experimental results; to explore the effect of varying the control parameters; to test new injection strategies; and to study the use of new materials. In addition, the results should be able to provide the starting point and the ``ground truth'' for reduced order (and simplified) models that could be used to simulate the fabrication of complete artifacts. The approach therefore plays a role comparable to direct numerical simulations in studies of turbulence (\cite{Pope2000, Sagaut2006}) and multiphase flows (\cite{TSZ:2011}). 

The development of a numerical framework like the one provide here should allow researchers  to focus on capturing the various physical processes that are important in FDM/FFF, unencumbered by the difficulty of solving the resulting equations, and to be able to obtain accurate solutions for nontrivial geometries that are similar to the structures encountered in practical applications.

\section{Acknowledgments}

Part of this research project was conducted using computational resources at the Maryland Advanced Research Computing Center (MARCC).

\bibliography{FDMBib}%,/Users/gretartryggvason/Desktop/BibtexFiles/GrandBib,/Users/gretartryggvason/Desktop/BibtexFiles/TryggvasonBib}

\end{document}